\documentclass[twoside]{article}
\usepackage{amsmath}
\usepackage{amssymb}
\usepackage{bm}
\usepackage{pifont}
\usepackage{qic,epsfig}

\def\re{\mathop{\bf Re}\nolimits}

\def\Tr{\mathop{\rm Tr}\nolimits}

\def\real{\mathbb{R}}

\newcommand{\defeq}{\stackrel{\rm def}{=}}

\def\max{\mathop{\rm max}}

\newcommand{\rhon}{\rho^{\otimes n}}
\newcommand{\sigman}{\sigma^{\otimes n}}

\def\rP{{\rm P}}

\def\bM{{\bm M}}

\def\mix{{\rm mix}}

\begin{document}
\setlength{\textheight}{8.0truein}    

\runninghead{Characterization of quantum analogues of relative entropy}
            {Masahito Hayashi}
\normalsize\textlineskip
\thispagestyle{empty}
\setcounter{page}{1}

\copyrightheading{0}{0}{2003}{000--000}

\vspace*{0.88truein}

\alphfootnote

\fpage{1}

\centerline{\bf
Characterization of several kinds}
\vspace*{0.035truein}
\centerline{\bf of quantum analogues of relative entropy}
\vspace*{0.37truein}
\centerline{\footnotesize
Masahito Hayashi}
\vspace*{0.015truein}
\centerline{\footnotesize\it 
ERATO Quantum Computation and Information Project, JST}
\baselineskip=10pt
\centerline{\footnotesize\it 
Hongo White Building, 5-28-3 Hongo, Bunkyo-ku, Tokyo 113-0033, Japan}
\centerline{\footnotesize\it Superrobust Computation Project}
\centerline{\footnotesize\it Information Science and Technology Strategic Core 
(21st Century COE by MEXT)}
\centerline{\footnotesize\it 
Graduate School of Information Science and Technology,
The University of Tokyo}
\centerline{\footnotesize\it 
7-3-1 Hongo, Bunkyo-ku, Tokyo 113-0033, Japan}

\vspace*{0.225truein}
\publisher{(received date)}{(revised date)}

\vspace*{0.21truein}

\abstracts{
Quantum relative entropy 
$D(\rho\|\sigma)\defeq\Tr \rho (\log \rho- \log \sigma)$
plays an important role in quantum information and 
related fields.
However, there are many quantum analogues of relative entropy.
In this paper, we characterize these analogues from information geometrical 
viewpoint. We also consider the naturalness of quantum relative entropy 
among these analogues.
}{}{}

\vspace*{10pt}

\keywords{Autoparallel curve, Divergence,  Fisher information,
Monotonicity, Additivity}
\vspace*{3pt}
\communicate{to be filled by the Editorial}

\vspace*{1pt}\textlineskip    

\section{Introduction}
In the quantum information theory, 
we usually focus on the quantum relative entropy
$D(\rho\|\sigma)\defeq\Tr \rho (\log \rho- \log \sigma)$
as a quantum analogue of relative entropy
(divergence).
However, there are many kinds of quantum analogues of relative entropy.
Some of them have been discussed from the viewpoint of operator 
algebra \cite{HP,HP2}.
In the classical information geometry,
the divergence can be defined by using the integral 
along the autoparallel curve.
Since the geometrical approach in classical information systems
is very attractive,
excellent insights for quantum information system 
can be expected through the consideration from 
geometrical viewpoints.

By extending this definition to the quantum system,
Nagaoka \cite{Nag94,Na-pra}
defined quantum analogues of divergence
based on the integral along the parallel translation.
$e$-parallel translation and
$m$-parallel translation are known 
as most popular parallel translations in the quantum system
as well as in the classical system.
These divergences are called 
$e$-path-divergence and $e$-path-divergence, respectively.
In particular, 
In the classical system, the path-divergences of both 
translations give the usual relative entropy.
On the other hand,
Fisher information is unique in the classical system.
However, it is not unique in the quantum system.
Petz\cite{Petz3} completely characterized 
its quantum analogues.
As famous examples, 
SLD Fisher information, RLD Fisher information,
and Bogoljubov Fisher information are known\cite{Helstrom:1967,Petz3,Petz4,Na,HolP}.
Nagaoka showed that 
the quantum path-divergence 
concerning $e$ ($m$)-parallel translation
coincides with 
the quantum relative entropy $D(\rho\|\sigma)$
when the quantum Fisher information of interest is
Bogoljubov Fisher information\cite{Nag94,Na-pra}.
He also calculated 
the quantum path-divergence 
with the SLD Fisher information concerning 
$e$-parallel translation\cite{Nag94}.

In this paper, we calculate
the quantum path-divergence other than the above cases.
Then, we succeeded in relating 
information geometrical path-divergence and
an operator-algebraic divergence
$\overline{D}(\rho\|\sigma)
= \Tr \rho \log (\rho^{\frac{1}{2}}\sigma^{-1}\rho^{\frac{1}{2}})$,
was introduced through operator-algebraic context
by Belavkin and Staszewski \cite{BS}.
Further, we proved 
the additivity of quantum path-divergence 
defined by $e$-parallel translation,
and the monotonicity of quantum path-divergence 
defined by $m$-parallel translation.
These two parallel translations are
the dual parallel translations of each other.
Since these two properties are fundamental, 
they are expected to be applied in the research field of quantum information.

In the classical system, the divergence also can be defined from 
a convex function.
Hence, divergence is closely related to convex analysis.
Amari \& Nagaoka \cite{Na} showed that 
only Bogoljubov Fisher information has zero-torsion.
That is, the geometry of Bogoljubov inner product has
the dual flat structure.
They also proved 
the equivalence of the following two conditions.
1) The path-divergences of dual parallel translations
can be given from potential function.
2) The dual parallel translation has the dual flat structure.
Hence, in the quantum case, we can conclude that 
only path-divergences of Bogoljubov Fisher information is given by a potential function.
This result indicates that
the geometry of Bogoljubov Fisher information
is closely related to optimization problem in quantum system.
In fact, in their proof, the calculations concerning
Christoffel symbols were essentially used.
However, many quantum information scientists are not familiar to 
such analysis.
In this paper, we give another proof of this argument without any 
use of Christoffel symbols.
This paper can be expected to be 
a good guidance for quantum information geometry
for quantum information scientist.

This paper is organized as follows.
In section \ref{2}, we review the information geometrical characterization 
of divergence $D(p\|q)$ in the classical system.
we also review how the divergence can be defined by
the convex function in the classical system.
In section \ref{3}, we give a review of inner product in quantum systems,
which is a fundamental of quantum information geometry.
In section \ref{4}, two kinds of autoparallel translations and 
autoparallel curves are reviewed.
In section \ref{5}, 
we treat quantum analogues of relative entropy 
from the operator-algebraic viewpoint.
In section \ref{6}, 
we examine quantum path-divergences based on $e$-autoparallel translation,
and consider their properties.
In section \ref{7}, 
we examine quantum path-divergences based on $m$-autoparallel translation,
and consider their properties.
In particular, the relation between 
an operator-algebraic divergence and quantum path-divergences 
based on $e$ ($m$)-autoparallel translation
are derived in section \ref{6} (\ref{7}).

\section{Divergence in Classical Systems}\label{2}
First, we review the information geometrical characterization 
of divergence $D(p\|q)$ in the classical system \cite{Na}.
Let $p(\omega)$ be a probability distribution, and $X(\omega)$ be 
a random variable.
When the family $\{p_\theta | \theta \in \Theta\}$ has the form
\begin{align}
p_\theta(\omega)& =
p(\omega) e^{\theta X(\omega)-\mu(\theta)} 
\label{4-10-11}\\
\mu(\theta) &\defeq
\log \sum_{\omega} p(\omega) e^{\theta X(\omega)}
\label{5-1-5},
\end{align}
the logarithmic derivative at respective points
equals the logarithmic derivative at a fixed point
with the addition of a constant.
In this case, the family is called an exponential family,
and $\mu(\theta)$ is called the moment function of $X$.
In particular, since the logarithmic derivative is closely related 
to exponential families,
it is often called the exponential ($e$) representation of the derivative.
Therefore, we use the superscript $(e)$ in the inner product
$\langle~, ~\rangle^{(e)}_p$.
The function $\mu(\theta)$ is often called 
a potential function
in the context of information geometry.
Since the second derivative $\mu''(\theta)$ is 
the Fisher information $J_\theta \ge 0$,
the moment function $\mu(\theta)$ is 
a convex function.
Therefore, the first derivative $\mu'(\theta)
=\sum_\omega p_\theta (\omega) X(\omega)$ is 
monotone increasing.
That is, 
we may regard it as another parameter identifying 
the distribution $p_\theta$, and denote it by $\eta$.
The original parameter $\theta$ is called a natural parameter
and the other parameter $\eta$ 
is an expectation parameter.
For example, 
in the binomial distribution,
the parameterization $p_\theta(0)= 1/(1+e^{\theta})$, 
$p_\theta(1)= e^{\theta}/(1+e^{\theta})$ 
is the natural parameter, and
the parameterization $p_\eta(0)=\eta$, $p_\eta(1)=1-\eta$ 
is the expectation parameter.
Hence, the binomial distribution is an exponential family.

Further, let $X_1(\omega), \ldots, X_k(\omega)$ be $k$ 
random variables.
We can define a $k$-parameter exponential family
\begin{align}
p_\theta(\omega)
&\defeq 
p(\omega) e^{\sum_i \theta^i X_i(\omega)- \mu(\theta)} , \nonumber \\
\mu(\theta)
&\defeq 
\log \sum_\omega p(\omega)
e^{\sum_i \theta^i X_i(\omega)} .\label{7-1-1}
\end{align}
The parameters $\theta^i$ are natural parameters,
and the other parameters
$\eta_i \defeq
\frac{\partial \mu}{\partial \theta^i}
=\sum_\omega p_\theta(\omega)X_i(\omega)$
are expectation parameters.
Since the second derivative $\frac{\partial^2 \mu(\theta)}
{\partial \theta^j \partial \theta^i}$ is equal to 
the Fisher Information matrix $J_{\theta:i,j}$,
the moment function $\mu(\theta)$ is
a convex function.

Let $\mu(\theta)$ be a twice-differentiable 
and strictly convex function defined 
on a subset of 
the $d$-dimensional real vector space $\real^d$.
The divergence concerning the convex function $\mu$ 
is defined by
\begin{align}
D^\mu(\bar{\theta}\|\theta) 
&\defeq
\sum_i \eta_i(\bar{\theta})
(\bar{\theta}^i-\theta^i) - 
\mu(\bar{\theta})+\mu(\theta), \nonumber \\
\eta_i(\theta)
&\defeq \frac{\partial \mu}{\partial \theta^i}(\theta)\label{7-3-9}.
\end{align}
This quantity has the following two characterizations:
\begin{align}
D^\mu(\bar{\theta}\|\theta) 
=&
\max_{\tilde{\theta}}
\frac{\partial \mu}{\partial \theta^i}(\bar{\theta})
(\tilde{\theta}^i-\theta^i) 
- \mu(\tilde{\theta})+\mu(\theta) \nonumber \\
=&
\int_0^1
\sum_{i,j} 
(\bar{\theta}^i-\theta^i )
(\bar{\theta}^j-\theta^j )
\frac{\partial^2 \mu}{\partial \theta^i\partial \theta^j}
(\theta+ (\bar{\theta}-\theta)t)
t d t.\label{7-4-4}
\end{align}
In the one-parameter case,
we obtain
\begin{align}
&D^\mu(\bar{\theta}\|\theta)
=
\mu'(\bar{\theta})(\bar{\theta}-\theta)
- \mu(\bar{\theta})+\mu(\theta)\nonumber\\
=&
\max_{\tilde{\theta}}
\mu'(\bar{\theta})(\tilde{\theta}-\theta)
- \mu(\tilde{\theta})+\mu(\theta)
=
\int_\theta^{\bar{\theta}}
\mu''(\tilde{\theta})
(\tilde{\theta}- \theta)
d \tilde{\theta}.\label{6-30-3}
\end{align}

Since the function $\mu$ is 
strictly convex, 
the correspondence $\theta^i \leftrightarrow 
\eta_i = \frac{\partial \mu}{\partial \theta^i}$ 
is one-to-one.
Hence, 
the divergence 
$D^\mu(\bar{\theta}\|\theta)$ can be expressed
with the parameter $\eta$.
For this purpose, 
we define the Legendre transform $\nu$ of $\mu$
\begin{align}
\nu(\eta)\defeq
\max_{\tilde{\theta}}
\sum_i \eta_i \tilde{\theta}^i - \mu(\tilde{\theta}).
\end{align}
Then, the function $\nu$ is a convex function,
and we can recover the function $\mu$ and $\theta$ as
\begin{align*}
\mu(\theta)
 =\max_{\tilde{\eta}}
\sum_i \theta_i \tilde{\eta}^i - \nu(\tilde{\eta})
 , \quad 
\theta^i
= \frac{\partial \nu}{\partial \eta_i}.
\end{align*}
The second derivative matrix
$\frac{\partial^2 \nu}{\partial \eta_i \partial \eta_j}$
of $\nu$  is equal to the inverse of 
the matrix
$\frac{\partial^2 \mu}{\partial \theta^i \partial \theta^j}$.

In particular,
when $\eta_i = \frac{\partial \mu}{\partial \theta^i}(\theta)$,
\begin{align}
\nu(\eta) &= \sum_i \eta_i {\theta}^i - \mu({\theta}) 
= D^\mu(\theta\|0)- \mu(0) ,\label{7-1-2}\\
\mu(\theta)&= \sum_i \theta_i {\eta}^i - \nu({\eta})
=D^\nu(\eta\|0)-\nu(0).
\end{align}
Using this relation, 
we can characterize
the divergence concerning the convex function $\mu$ 
by 
the divergence concerning the convex function $\nu$ as
\begin{align}
D^\mu(\bar{\theta}\|\theta)
= D^\nu(\eta\|\bar{\eta})
= 
\sum_i \theta^i(\eta_i-\bar{\eta}_i) 
- \nu(\eta)+\nu(\bar{\eta}). 
\label{6-30-1}
\end{align}

Now, we apply the discussion about the divergence to
a multi-parametric exponential family $\{p_\theta| \theta \in \real \}$
defined in (\ref{7-1-1}) \cite{Na}.
Then, 
\begin{align*}
 D(p_{\bar{\theta}}\|p_\theta)
= D^\mu(\bar{\theta}\|\theta) 
= \sum_i \eta_i(\bar{\theta})(\bar{\theta}^i-\theta^i) - 
\mu(\bar{\theta})+\mu(\theta).
\end{align*}
In particular, 
applying (\ref{6-30-3}) to a one-parameter 
exponential family (\ref{4-10-11}),
we have
\begin{align}
&D(p_{\bar{\theta}}\|p_{\theta}) 
 = D(p_{\eta(\theta)+\epsilon}\|p_{\eta(\theta)})
= (\bar{\theta}-\theta)\eta(\bar{\theta})
- \mu(\bar{\theta})+\mu(\theta)\nonumber \\
= &\int_\theta^{\bar{\theta}} 
J_{\tilde{\theta}}(\tilde{\theta} - \theta)
d \tilde{\theta}
= \max_{\tilde{\theta} :\tilde{\theta} \ge \theta} 
(\tilde{\theta}-\theta) (\eta(\theta)+ \epsilon) 
- \mu(\tilde{\theta})+ \mu(\theta)\label{5-1-8-1}.
\end{align}

In the following, we consider the case where
$p$ is the uniform distribution $p_{\mix}$.
Let the random variables $X_1(\omega), \ldots, X_k(\omega)$ 
be a CONS of the space of random variables with expectation $0$ 
under the uniform distribution $p_{\mix}$,
and $Y^1(\omega), \ldots, Y^k(\omega)$
be its dual basis 
satisfying 
$\sum_\omega Y^i(\omega)X_j(\omega)= \delta^i_j$.
Then, any distribution can be parameterized by the 
expectation parameter as
\begin{align*}
p_{\eta(\theta)}(\omega) 
= p_{\mix}(\omega) +\sum_{i} \eta_i(\theta) Y^i(\omega).
\end{align*}
From (\ref{6-30-1}) and (\ref{7-1-2}),
\begin{align*}
D(p_{\bar{\eta}}\|p_{\eta})
&=
D^\nu(\eta\|\bar{\eta})
=
\frac{\partial \nu}{\partial \eta_i}
(\eta_i - \bar{\eta}_i)
-\nu({\eta})+ \nu(\bar{\eta})\\
\nu(\eta)
&=D(p_\eta\|p_{\mix})= - H(p_\eta)+ H(p_{\mix})
\end{align*}
because $\mu(0)=0$.
The second derivative matrix of $\nu$ is
the inverse of the second derivative matrix of $\mu$, {\it i.e.},
the Fisher information matrix concerning the natural parameter $\theta$.
That is,
the second derivative matrix of $\nu$ coincides with 
the Fisher information matrix concerning the expectation parameter $\eta$.
Hence, applying (\ref{6-30-3}) to
the subspace $\{(1-t) p+ t q|0 \le t \le1 \}$,
we have
\begin{align}
D(p\|q)
=\int_0^1 J_t t dt ,\label{7-2-7}
\end{align}
where $J_t$ is the Fisher information concerning
the parameter $t$.

\section{Inner Products in Quantum Systems}\label{3}
In this section, 
in order to define the quantum analogues of divergence,
we define as inner products in quantum systems.
There are at least three possible ways of 
defining the product corresponding to $X\rho$: 
\begin{align}
E_{\rho,s}(X) & \defeq X \circ \rho \defeq 
\frac{1}{2}\left(\rho X + X \rho \right) \label{6-20-1},\\
E_{\rho,b}(X) & \defeq 
\int_0^1 \rho^{\lambda}  X \rho^{1-\lambda} \,d \lambda , \nonumber\\
E_{\rho,r}(X) & \defeq \rho X .\label{6-20-3}
\end{align}
Here, $X$ is not necessarily Hermitian.
These extensions are unified in the general form \cite{Petz4}
\begin{align}
E_{\rho,p}(X) & \defeq 
\int_0^1 E_{\rho,\lambda}(X) p(d \lambda) ,
\label{6-20-4-1} \\
E_{\rho,\lambda}(X) & \defeq 
\rho^{\lambda}  X \rho^{1-\lambda} ,
\label{7-2-1}
\end{align}
where $p$ is an arbitrary probability distribution
on $[0,1]$.
The case (\ref{6-20-1}) corresponds to 
the case (\ref{6-20-4-1}) with $p(1)=p(0)=1/2$,
and the case  (\ref{6-20-3}) does to
the case (\ref{6-20-4-1}) with $p(1)=1$.
In particular,
the map $E_{\rho,x}$ is symmetric,
when $E_{\rho,x}(X)$ is Hermitian if and only if 
$X$ is Hermitian.
Hence, when the distribution $p$ is symmetric,
{\it i.e.}, $p(x)=p(1-x)$,
the map $E_{\rho,p}$ is symmetric.
When $\rho \,> 0$, these maps possess inverses.

Accordingly, we may define these types of inner products
\begin{align*}
\langle Y, X \rangle_{\rho,x}^{(e)} \defeq
\Tr Y^* E_{\rho,x}(X) \quad x= s,b,r ,\lambda, p .
\end{align*}
If $X,Y,\rho$ all commute, 
these have the same value. 
These are called the SLD, Bogoljubov\footnote{The Bogoljubov inner product is also 
called the canonical correlation in statistical mechanics.  In 
linear response theory, it is often used to give an approximate correlation between two 
different physical quantities.}, RLD, $\lambda$, and $p$ inner 
products\cite{Helstrom:1967,Petz3,Petz4,Na,HolP},
respectively (reasons for this will 
be given in the next section).  These inner products
are positive semi-definite and Hermitian,
{\it i.e.},
\begin{align*}
\left(\|X\|_{\rho,x}^{(e)}\right)^2 
\defeq \langle X, X \rangle_{\rho,x}^{(e)} \ge 0,~
\langle Y, X \rangle_{\rho,x}^{(e)} =
(\langle X, Y \rangle_{\rho,x}^{(e)} )^*  .
\end{align*}
A dual inner product may be defined $\langle A, B 
\rangle_{\rho,x}^{(m)}
\defeq \Tr (E_{\rho,x}^{-
1}(A))^* B$ with respect to the correspondence 
$A= E_{\rho,x}(X)$.  Denote the norm of these inner products as
$\left(
\|A\|_{\rho,x}^{(m)} 
\right)^2  \defeq \langle A, A \rangle_{\rho,x}^{(m)}$.  
Hence, the inner product
$\langle A, B \rangle^{(m)}_{\rho,x}$ is
positive semi-definite and Hermitian.
Using this inner product, 
we define quantum analogues of Fisher information as
\begin{align*}
J_{\theta_0,x}\defeq 
\left(\left\|\frac{\,d \rho_\theta}{\,d \theta}(\theta_0)
\right\|_{\rho_{\theta_0},x}^{(m)}\right)^2 
\end{align*}
for a one-parameter family $\{\rho_\theta\}$
and $x= s,r,b,\lambda,p$.

\section{Autoparallel Curves in Quantum Systems}\label{4}
Next, we define parallel transport and autoparallel curves
in quantum systems according to Nagaoka \cite{Nag94} and 
Amari \& Nagaoka\cite{Na}.
To introduce the concept of a parallel transport, 
consider an infinitesimal displacement in 
a one-parameter quantum state family 
$\{\rho_\theta| \theta \in \real\}$.  
The difference between $\rho_{\theta+\epsilon}$ and $\rho_\theta$ 
approximately equals to $\frac{\,d \rho_\theta}{\,d 
\theta}(\theta) \epsilon$.  
Hence, the state  $\rho_{\theta+\epsilon}$ can be regarded
as the state transported from the state $\rho_\theta$ 
in the direction $\frac{\,d \rho_\theta}{\,d \theta}(\theta)$ 
by an amount $\epsilon$. 
However, if the state $\rho_{\theta+\epsilon}$ coincides precisely 
with
the state displaced from the state $\rho_\theta$ by $\epsilon$ 
in the direction of $\frac{\,d \rho_\theta}{\,d \theta}(\theta)$, 
the infinitesimal displacement at the intermediate states 
$\rho_{\theta+\epsilon'}$ ($0\,<\epsilon'\,< \epsilon$) 
must equal the 
infinitesimal displacement
$\frac{\,d \rho_\theta}{\,d \theta}(\theta)\Delta$ at $\theta$.  
Then, the problem is to ascertain
which infinitesimal displacement at the point $\theta+\epsilon'$ corresponds 
to the given infinitesimal displacement 
$\frac{\,d \rho_\theta}{\,d \theta}(\theta)\Delta$ 
at the initial point $\theta$.  
The rule for matching the 
infinitesimal displacement at one point 
to the infinitesimal displacement at another point 
is called parallel transport.  
The coefficient $\frac{\,d \rho_\theta}{\,d 
\theta}(\theta)$ of the infinitesimal displacement at $\theta$ 
is called the tangent vector, as it 
represents the slope of the tangent line of the state family $\{\rho_\theta|\theta\in \real \}$ at 
$\theta$.  Therefore, 
we can consider the parallel transport of a tangent vector instead 
of the parallel transport of an infinitesimal displacement.  

Commonly used parallel transports can be classified into 
those based on 
the $ m $ representation ($m$ parallel translation)
and those based on the $ e $ representation 
($e$ parallel translation).  
The $ m $ parallel translation $\Pi_{\rho_{\theta},\rho_{\theta'}}^{(m)}$
moves
the tangent vector at one point $\rho_{\theta}$
to the tangent vector with the same $ m $ representation
at another point $\rho_{\theta'}$.
On the other hand, the $ e $ parallel translation
$\Pi_{x,\rho_{\theta},\rho_{\theta'}}^{(e)}$
moves the tangent vector at one point $\rho_\theta$
with the $ e $ representation $ L $
to the tangent vector at another point $\rho_{\theta'}$
with the $ e $ representation $L - \Tr \rho_{\theta'} L$\cite{Na}.
Of course, this definition requires 
the coincidence between
the set of $ e $ representations at the point $\theta$ and 
that at another point $\theta'$.
Hence, this type of $e$ parallel translation
is defined only for the symmetric inner product 
$\langle X,Y\rangle^{(e)}_{\rho,x}$,
and its definition depends on the choice of the metric.
Indeed, the $ e $ parallel translation can be regarded as
the dual parallel translation of the $ m $ parallel translation
concerning the metric $\langle X,Y\rangle^{(e)}_{\rho,x}$
in the following sense:
\begin{align*}
\Tr X^* \Pi_{\rho_{\theta},\rho_{\theta'}}^{(m)}(A)=
\Tr \Pi_{x,\rho_{\theta'},\rho_{\theta}}^{(e)}(X)^* A,
\end{align*}
where $X$ is the $e$ representation of a tangent vector at 
$\rho_{\theta'}$
and $A$ is the $m$ representation of another tangent vector at 
$\rho_{\theta}$.

Further, a one-parameter quantum state 
family is called a geodesic or an 
autoparallel curve when 
the tangent vector ({\it i.e.} the derivative)
at each point 
is given as a parallel transport of a tangent vector at a fixed point.
Especially, the $e$ geodesic
is called a one-parameter exponential family.

For example, in an $e$ geodesic with respect to SLD
$\{\rho_\theta | \theta \in \real \}$,
any state $\rho_{\theta}$
coincides with the state transported from the state  $\rho_0$ along 
the autoparallel curve in the direction $L$ by an amount $\theta$,
where $L$ denotes the SLD $e$ representation of the derivative 
at $\rho_0$.
We shall henceforth denote the state as $\Pi^\theta_{L,s} \rho_0$.  
Similarly, $\Pi^\theta_{L,b} \rho_0$
denotes the state transported autoparallely
with respect to the Bogoljubov $e$ representation from $\rho_0 $ 
in the direction $L$ by an amount $ \theta$.

When the given metric is not symmetric,
the $e$ parallel translation
moves the tangent vector at one point $\theta$
under the $ e $ representation $\tilde{L}$
to the tangent vector at another point $\theta'$
with the $ e $ representation $\tilde{L}' - \Tr \rho_{\theta'} \tilde{L}'$
with the condition $\tilde{L} + \tilde{L}^* 
= \tilde{L}' + (\tilde{L}')^*$.
That is, we require the same Hermitian part in the $e$ representation.
Hence, 
the $ e $ parallel translation
$\Pi_{x,\rho_{\theta},\rho_{\theta'}}^{(e)}$
coincides with 
the $ e $ parallel translation
$\Pi_{s(x),\rho_{\theta},\rho_{\theta'}}^{(e)}$
with regard to its symmetrized inner product.
Therefore, we can define
the state 
transported from the state  $\rho_0$ along 
the autoparallel curve in the direction with the Hermitian part $L$
by an amount $\theta$
with respect to RLD ($\lambda$, $p$), 
and denote them by 
$\Pi^\theta_{L,r} \rho_0$
($\Pi^\theta_{L,\lambda} \rho_0$,
 $\Pi^\theta_{L,p} \rho_0$),
respectively.
However, only the 
SLD one-parameter exponential family
$\{\Pi^\theta_{L,s} \rho_0| s \in \real \}$ 
plays an important role in quantum estimation 
examined in the next section.  

\begin{lemma}\label{3-20t3}
$\Pi^\theta_{L,s} \sigma$, $\Pi^\theta_{L,b} \sigma$, 
$\Pi^\theta_{L,r} \sigma$ and $\Pi^\theta_{L,\frac{1}{2}} \sigma$ 
may 
be written in the following form\cite{Nagaoka:1989:2,Nag94,Na}:
\begin{align}
\Pi^\theta_{L,s} \sigma &=
e^{-\mu_s(\theta)}
e^{\frac{\theta}{2} L} 
\sigma 
e^{\frac{\theta}{2} L} ,
\label{3-19-5} \\
\Pi^\theta_{L,b} \sigma &=
e^{-\mu_b(\theta)}
e^{ \log \sigma + \theta L} \label{3-19-5.1},\\
\Pi^\theta_{L,r} \sigma &=
e^{-\mu_r(\theta)}
\sqrt{\sigma}
e^{ \theta L_r} 
\sqrt{\sigma}
\label{3-19-5.2},\\
\Pi^\theta_{L,\frac{1}{2}} 
\sigma &=
e^{-\mu_{\frac{1}{2}}(\theta)}
\sigma^{\frac{1}{4}}
e^{ \frac{\theta}{2} L_{\frac{1}{2}}}
\sigma^{\frac{1}{2}}
e^{ \frac{\theta}{2} L_{\frac{1}{2}}}
\sigma^{\frac{1}{4}}
\label{3-19-5.3},
\end{align}
where we choose Hermitian matrices $L_r$ and $L_{\frac{1}{2}}$
as 
$L=
\frac{1}{2}
(\sigma^{-\frac{1}{2}}
L_r \sigma^{\frac{1}{2}}+
\sigma^{\frac{1}{2}} 
L_r \sigma^{-\frac{1}{2}}
)
$ and 
$L=
\frac{1}{2}(\sigma^{-\frac{1}{4}}L_{\frac{1}{2}}
\sigma^{\frac{1}{4}}
+ \sigma^{\frac{1}{4}}L_{\frac{1}{2}}
\sigma^{-\frac{1}{4}})$, respectively, and
\begin{align}
\mu_s(\theta)
&\defeq  
\log \Tr e^{\frac{\theta}{2} L} 
\sigma e^{\frac{\theta}{2} L} \nonumber \\
\mu_b(\theta)
& \defeq
\log \Tr e^{ \log \sigma + \theta L},\label{9-18-6}
\\
\mu_r(\theta)
&\defeq  
\log \Tr \sqrt{\sigma}e^{ \theta L_r} \sqrt{\sigma},\nonumber \\
\mu_{1/2}(\theta)
&\defeq  
\log \Tr 
\sigma^{\frac{1}{4}}
e^{ \frac{\theta}{2} L_{\frac{1}{2}}} 
\sigma^{\frac{1}{2}}
e^{ \frac{\theta}{2} L_{\frac{1}{2}}}  
\sigma^{\frac{1}{4}}.
\nonumber
\end{align}
\end{lemma}

\proof{Taking the derivative of the RHS of 
(\ref{3-19-5}) and (\ref{3-19-5.1}), 
we see that the SLD (or Bogoljubov) $ e $ representation of the derivative 
at each point is equal to the parallel 
transported $ e $ representation of the derivative 
$ L$ at $\sigma$.  
In the RHS of (\ref{3-19-5.2}), 
the RLD $ e $ representation of the derivative 
at each point is equal to the parallel 
transported $ e $ representation of the derivative 
$\sqrt{\sigma}^{-1}L_r\sqrt{\sigma}$ at $\sigma$.  
Further, 
In the RHS (\ref{3-19-5.3}), 
the $\frac{1}{2}$ $ e $ representation of the derivative 
at each point is equal to the parallel 
transported $ e $ representation of the derivative 
$L_r$ at $\sigma$.  

Conversely, from 
the definition of  $\Pi^\theta_{L,x} \sigma$, we have
\begin{align*}
\frac{\,d \Pi^\theta_{L,x} \sigma}{\,d \theta}
=
E_{\rho_\theta,x}(L - \Tr L \rho_\theta), \quad 
x= s,r , \frac{1}{2}.
\end{align*}
Since this has only one variable, this is actually an 
ordinary differential equation.  
From the uniqueness of the solution of an ordinary differential 
equation, the only $\Pi^\theta_{L,x} \sigma$ satisfying 
$\Pi^0_{L,x} \sigma = \sigma$ is the one given above. 
Since any $e$ representation $\sigma$ has the form
$\sqrt{\sigma}^{-1}L \sqrt{\sigma}$ with a Hermitian matrix $L$,
we only discuss 
$\rho_\theta= \Pi^\theta_{\sqrt{\sigma}^{-1}L\sqrt{\sigma},r} 
\sigma$.
Taking its derivative, we have
\begin{align*}
\frac{\Pi^\theta_{\sqrt{\sigma}^{-1}L\sqrt{\sigma},r} \sigma}
{d \theta}=
\rho_\theta (\sqrt{\sigma}^{-1}L\sqrt{\sigma}-
\Tr \rho_\theta\sqrt{\sigma}^{-1}L\sqrt{\sigma}) .
\end{align*}
Similarly, from the uniqueness of the solution of 
an ordinary differential equation, 
only the state family (\ref{3-19-5.2}) satisfies 
this condition.}
\section{Non-Geometrical Characterization of Divergences 
in Quantum Systems}\label{5}
First, we briefly characterize quantum analogues of divergence
from the non-geometrical viewpoint.
A quantity $\tilde{D}(\rho\|\sigma)$
can be regarded as a quantum version of divergence
if any commutative states $\rho$ and $\sigma$ satisfy
\begin{align}
\tilde{D}(\rho\|\sigma)
= D(p\|\bar{p}),\label{7-5-4}
\end{align}
where $p$ and $\bar{p}$ is the probability distribution
consisting of the eigenvalues of $\rho$ and $\sigma$.
If a relative entropy $\tilde{D}(\rho\|\sigma)$
satisfies the monotonicity for a POVM $\bM= \{M_i\}$:
\begin{align}
\tilde{D}(\rho\|\sigma)
\ge D(\rP_\rho^{\bM}\|\rP_\sigma^{\bM})
,~
\rP_\rho^{\bM}(i)\defeq \Tr \rho M_i
\label{7-5-2}
\end{align}
and the additivity
\begin{align}
\tilde{D}(\rho_1 \otimes \rho_2\|
\sigma_1 \otimes \sigma_2 )=
\tilde{D}(\rho_1\|\sigma_1)
+\tilde{D}(\rho_2\|\sigma_2),\label{7-5-10}
\end{align}
then Hiai \& Petz \cite{HP}'s result yields the relation
\begin{align}
 \tilde{D}(\rho\|\sigma) 
= \lim \frac{\tilde{D}(\rhon\|\sigman) }{n} 
\ge 
\lim \sup_{\bM} \frac{D(\rP_{\rhon}^{\bM}\|\rP_{\sigman}^{\bM})}{n} 
= D(\rho\|\sigma)\label{7-5-1}.
\end{align}
That is, the quantum relative entropy $D(\rho\|\sigma)$
is the minimum quantum analogue of 
relative entropy with the monotonicity for measurement and the additivity.

Further, Hiai \& Petz \cite{HP} 
showed the inequality
\begin{align}
D(\rho\|\sigma)
 \le \overline{D}(\rho\|\sigma)
 \defeq 
 \Tr \rho \log (\rho^{\frac{1}{2}}\sigma^{-1}\rho^{\frac{1}{2}})
 . \label{9-26-22}
\end{align}

\section{Quantum Path-divergences Based on $e$-Parallel Translation}\label{6}
Now, using the concept of 
the exponential family, we extend the path-divergence based on 
the first equation in (\ref{5-1-8-1}).
For any two states $\rho$ and $\sigma$,
we choose the Hermitian matrix $L$ such that the exponential family 
$\{\Pi^\theta_{L,x} \sigma \}_{\theta \in [0,1]}$
concerning the inner product $J_{\theta,x}$
satisfies 
\begin{align}
\Pi^{1}_{L,x} \sigma= \rho.\label{7-2-5}
\end{align}
Then, we define the $x$-$e$-divergence as follows:
\begin{align}
D^{(e)}_x(\rho\|\sigma)
= \int_0^1 J_{\theta,x} \theta d \theta,
\end{align}
where $J_{\theta,x}$ is the Fisher information concerning
the exponential family $\Pi^\theta_{L,x} \sigma$.
Since $\Pi^\theta_{L^1 \otimes I + I \otimes L^2,x} 
(\sigma_1 \otimes \sigma_2) $ equals
$(\Pi^\theta_{L^1,x} \sigma_1)\otimes 
(\Pi^\theta_{L^2,x} \sigma_2) $,
\begin{align}
D^{(e)}_x(\rho_1\otimes \rho_2\|\sigma_1\otimes \sigma_2)=
D^{(e)}_x(\rho_1\|\sigma_1)+
D^{(e)}_x(\rho_2\|\sigma_2),\label{7-5-11}
\end{align}
{\it i.e.}, the $e$-divergence 
satisfies the additivity for any inner product.
\begin{theorem}
When 
\begin{align}
L= 
\left\{
\begin{array}{ll}
2\log \sigma^{-\frac{1}{2}}
(\sigma^{\frac{1}{2}}\rho\sigma^{\frac{1}{2}})^{\frac{1}{2}}
\sigma^{-\frac{1}{2}} & \hbox{ for }x= s\\
\log \rho -\log \sigma & \hbox{ for }x= b\\
\frac{1}{2}\bigl[
\sigma^{-\frac{1}{2}}
\log (\sigma^{-\frac{1}{2}}\rho\sigma^{-\frac{1}{2}})
\sigma^{\frac{1}{2}} & \hbox{ for }x= r\\
\quad + \sigma^{\frac{1}{2}}
\log (\sigma^{-\frac{1}{2}}\rho\sigma^{-\frac{1}{2}})
\sigma^{-\frac{1}{2}}\bigr] & \\
\sigma^{-\frac{1}{4}}
\log (\sigma^{-\frac{1}{4}}\rho^{\frac{1}{2}}\sigma^{-\frac{1}{4}})
\sigma^{\frac{1}{4}} &\hbox{ for }x= \frac{1}{2}\\
\quad + \sigma^{\frac{1}{4}}
\log (\sigma^{-\frac{1}{4}}\rho^{\frac{1}{2}}\sigma^{-\frac{1}{4}})
\sigma^{-\frac{1}{4}}, & 
\end{array}
\right.\label{7-3-1}
\end{align}
the condition (\ref{7-2-5}) holds.
Hence, we obtain
\begin{align}
D^{(e)}_s(\rho\|\sigma)&=
2 \Tr \rho \log 
\sigma^{-\frac{1}{2}}
(\sigma^{\frac{1}{2}}\rho\sigma^{\frac{1}{2}})^{\frac{1}{2}}
\sigma^{-\frac{1}{2}} \label{7-3-2.s}\\
D^{(e)}_b(\rho\|\sigma)&=
\Tr \rho (\log \rho -\log \sigma) =D(\rho\|\sigma)
\label{7-3-2.b}\\
D^{(e)}_r(\rho\|\sigma)&=
\Tr \rho 
\log (\rho^{\frac{1}{2}}
\sigma^{-1}\rho^{\frac{1}{2}})
=\overline{D}(\rho\|\sigma) 
\label{7-3-2.r}\\
D^{(e)}_{\frac{1}{2}}(\rho\|\sigma)&=
2 \Tr 
(\sigma^{\frac{1}{4}}
\rho^{\frac{1}{2}}
\sigma^{\frac{1}{4}})
(\sigma^{-\frac{1}{4}}\rho^{\frac{1}{2}}\sigma^{-\frac{1}{4}}) 
\log (\sigma^{-\frac{1}{4}}\rho^{\frac{1}{2}}\sigma^{-\frac{1}{4}}).
\label{7-3-2.1/2}
\end{align}
\end{theorem}
Nagaoka \cite{Nag94} obtained the above results for $x=s,b$.

\proof{When we substitute (\ref{7-3-1}) into $L$,
condition (\ref{7-2-5}) can be checked by 
using Lemma \ref{3-20t3}.
In this case, 
$L_r=
\log (\sigma^{-\frac{1}{2}}\rho\sigma^{-\frac{1}{2}})$,
$L_{\frac{1}{2}}=
2\log(\sigma^{-\frac{1}{4}}\rho^{\frac{1}{2}}\sigma^{-\frac{1}{4}})$,
and we can show that
\begin{align}
\frac{d^2 \mu_x(\theta)}{d \theta^2}
= J_{\theta,x}. 
\end{align}
Hence, from a discussion similar 
to (\ref{5-1-8-1}),
we can prove that
\begin{align}
 D_x^{(e)}(\rho\|\sigma)
= \left.\frac{d \mu_x(\theta)}{d \theta}
\right|_{\theta=1}(1-0)
- \mu_x(1)+\mu_x(0) 
= \left.\frac{d \mu_x(\theta)}{d \theta}
\right|_{\theta=1},\label{7-3-8}
\end{align}
where $\mu_x(\theta)$ is defined in Theorem \ref{3-20t3}.
Using this relation, we can check (\ref{7-3-2.s}),
(\ref{7-3-2.b}), and (\ref{7-3-2.1/2}).
Concerning (\ref{7-3-2.r}),
we obtain 
\begin{align*}
D^{(e)}_r(\rho\|\sigma)=
\Tr \sigma
\sigma^{-\frac{1}{2}}\rho\sigma^{-\frac{1}{2}}
\log (\sigma^{-\frac{1}{2}}\rho\sigma^{-\frac{1}{2}})
=
\Tr \rho 
\log (\rho^{\frac{1}{2}}\sigma^{-1}\rho^{\frac{1}{2}}),
\end{align*}
where the last equation follows from 
the equation with $AU= \sigma^{-\frac{1}{2}}\rho^{\frac{1}{2}}$
($A$ is Herimitain and $U$ is unitary):
\begin{align*}
AU U^* A \log (AU U^* A )
= AU \log (U^* AAU)  U^* A .
\end{align*}
}
Now, we compare these quantum analogues of relative entropy given in 
(\ref{7-3-2.s})--(\ref{7-3-2.1/2}).
As is easily checked, these satisfy the condition (\ref{7-5-4}) for 
quantum analogues of relative entropy.
Let $\bM$ be a measurement corresponding to the spectral decomposition of 
$\sigma^{-1/2} (\sigma^{1/2}\rho\sigma^{1/2})^{1/2}\sigma^{-1/2}$.
This PVM $\bM$ satisfies
that $D_s^{(e)}
(\rho\|\sigma)= D(\rP_\rho^{\bM}\|\rP_\sigma^{\bM})$.
Thus,
from the monotonicity for measurement 
concerning the quantum relative entropy $D(\rho\|\sigma)$,
\begin{align}
D(\rho\|\sigma)\ge D^{(e)}_s(\rho\|\sigma)
=
2 \Tr \rho \log 
\sigma^{-\frac{1}{2}}
(\sigma^{\frac{1}{2}}\rho\sigma^{\frac{1}{2}})^{\frac{1}{2}}
\sigma^{-\frac{1}{2}} .\label{7-5-6}
\end{align}
From (\ref{9-26-22}),
\begin{align}
D(\rho\|\sigma)\le D^{(e)}_{r}(\rho\|\sigma)
=\Tr \rho 
\log (\rho^{\frac{1}{2}}\sigma^{-1}\rho^{\frac{1}{2}})
.\label{7-5-7}
\end{align}
Hence,
from the inequality (\ref{7-5-1})
and the additivity (\ref{7-5-11}), 
$D^{(e)}_s(\rho\|\sigma)$
and $D^{(e)}_r(\rho\|\sigma)$ do not satisfy
the monotonicity even for measurements
because the equality in (\ref{7-5-6}) and (\ref{7-5-7}) does not 
always hold.

\section{Quantum Path-divergences Based on $m$-Parallel Translation}\label{7}
Further, we can extend the path-divergence based on 
the equation (\ref{7-2-7}).
For any two states $\rho$ and $\sigma$,
the family $\{(1-t)\rho + t\sigma | 0 \le t \le 1\}$
is the $m$ geodesic joining $\rho$ and $\sigma$.
Hence, as an extension of (\ref{7-2-7}),
we can define the $x$-$m$ divergence as 
\begin{align}
D^{(m)}_x(\rho\|\sigma)\defeq
\int_0^1 J_{t,x} t d t.
\end{align}
Since the family $\{(1-t)\kappa(\rho) + t\kappa(\sigma) | 0 \le t \le 1\}$
is the $m$ geodesic joining $\kappa(\rho)$ and $\kappa(\sigma)$
for any TP-CP map $\kappa$,
we have
\begin{align}
D^{(m)}_x(\rho\|\sigma)
\ge D^{(m)}_x(\kappa(\rho)\|\kappa(\sigma)),\label{7-5-14}
\end{align}
{\it i.e.}, the $m$ divergence satisfies the 
monotonicity.
Since the RLD is the largest inner product,
\begin{align}
D^{(m)}_r(\rho\|\sigma)
\ge D^{(m)}_x(\rho\|\sigma).\label{7-5-19}
\end{align}
We can calculate 
the $m$ divergence as
\begin{align}
D^{(m)}_b(\rho\|\sigma)&=
\Tr \rho (\log \rho -\log \sigma) =D(\rho\|\sigma)
\label{7-4-7.b}\\
D^{(m)}_r(\rho\|\sigma)&=
\Tr \rho 
\log (\sqrt{\rho}\sigma^{-1}\sqrt{\rho})
= \overline{D}(\rho\|\sigma)
.
\label{7-4-7.r}
\end{align}
In fact, The Bogoljubov case (\ref{7-4-7.b}) has been obtained by 
Nagaoka \cite{Na-pra},
and follows from Theorem \ref{7-4-5}.
Hence, $\Tr \rho \log (\sqrt{\rho}\sigma^{-1}\sqrt{\rho})= 
D^{(m)}_r(\rho\|\sigma)$ satisfies the monotonicity for 
TP-CP maps. Also, from (\ref{7-5-19}), we obtain  $\Tr \rho 
\log (\sqrt{\rho}\sigma^{-1}\sqrt{\rho})
\ge D(\rho\|\sigma)$\cite{HP}.

Further, all of $x$-$m$ divergences do not necessarily satisfy
the additivity (\ref{7-5-11}).
At least, 
when the inner product $J_{x,\theta}$ is
smaller than the Bogoljubov inner product $J_{b,\theta}$,
{\it i.e.},
$J_{\theta,x} \le J_{\theta,b}$,
we have $D(\rho\|\sigma)\ge D^{(m)}_x(\rho\|\sigma)$.
From (\ref{7-5-1}) and the monotonicity (\ref{7-5-14}),
$D^{(m)}_x(\rho\|\sigma)$ does not satisfy
the additivity (\ref{7-5-11}).
For example, SLD $m$ divergence does not satisfy
the additivity (\ref{7-5-11}).

We can now verify whether it is possible in two-parameter state 
families to have states that are $ e $ autoparallel 
transported in the direction of $L_1$ by $\theta^1$,
and in the direction $L_2$ by $\theta^2$.  
In order to define such a state, 
we require that the state that is $ e $ autoparallel transported 
first in the $L_1$ direction by $\theta^1$ from $\rho_0$, then 
further $ e $ autoparallel transported in the $L_2$ direction by 
$\theta^2$ coincides with the state that is 
$ e $ autoparallel transported in the $L_2$ direction by 
$\theta^2$ from $\rho_0$, then $ e $ autoparallel transported 
in the $L_1$ direction by $\theta^1$.  
That is, if such a state would be defined, the relation
\begin{align}
\Pi^{\theta^2}_{L_2,x} \Pi^{\theta^1}_{L_1,x} \sigma
=
\Pi^{\theta^1}_{L_1,x} \Pi^{\theta^2}_{L_2,x} \sigma 
\label{7-3-10}
\end{align}
should hold.
Concerning this condition,
we have the following theorem.
\begin{theorem}\label{7-4-5}
The following conditions for the inner product $J_{\theta,x}$
are equivalent
\begin{enumerate}
\item[\ding{192}] $J_{\theta,x}$ is the Bogoljubov inner product,
{\it i.e.}, $x=b$.
\item[\ding{193}] 
The condition (\ref{7-3-10}) holds for any 
two Hermitian matrices $L_1$ and $L_2$
and any state $\rho_0$.
\item[\ding{194}] 
$D^{(e)}_x(\rho_{\bar{\theta}}\|\rho_\theta)
= D^\mu(\bar{\theta}\|\theta)$. 
\item[\ding{195}] 
$D^{(e)}_x(\rho\|\sigma)
=D(\rho\|\sigma)$.
\item[\ding{196}] 
$D^{(m)}_x(\rho_{\bar{\eta}}\|\rho_\eta)
= D^\nu(\eta\|\bar{\eta})$.
\item[\ding{197}] 
$D^{(m)}_x(\rho\|\sigma)=D(\rho\|\sigma)$.
\end{enumerate}
Here, 
the convex functions $\mu(\theta)$, $\nu(\eta)$
and the states $\rho_\theta$, $\rho_\eta$ 
are defined by
\begin{align}
\rho_\theta \defeq &\exp(\sum_i \theta^i X_i-\mu(\theta)),\nonumber \\
\mu(\theta)\defeq &\log \Tr \exp(\sum_i \theta^i X_i),\label{7-3-4}\\
\rho_\eta \defeq &\rho_{\mix}+ \sum_j \eta_j Y^j, \nonumber \\
\nu(\eta)\defeq & D_x^{(m)}(\rho_0\|\rho_\eta)
= -H(\rho_\eta)+H(\rho_{\mix}),
\nonumber
\end{align}
where 
$X_1, \ldots, X_k$ is a basis of the set of traceless 
Hermitian matrices,
and $Y^1, \ldots, Y^k$ is its dual basis.
\end{theorem}
This theorem implies that only the quantum path-divergence based on 
the Bogoljubov Fisher information can be characterized by 
the convex function among quantum path-divergence based on 
$m$-parallel translation.

\proof{First, we prove that \ding{192}$\Rightarrow$\ding{193}.
Theorem \ref{3-20t3} guarantees that
Bogoljubov $ e $ autoparallel transport satisfies
\begin{align*}
\Pi^{\theta^2}_{L_2,b} \Pi^{\theta^1}_{L_1,b} \rho
=
\Pi^{\theta^1}_{L_1,b} \Pi^{\theta^2}_{L_2,b} \rho 
=
e^{-\mu_b(\theta^1,\theta^2)}
e^{ \log \rho + \theta^1 L_1+\theta^2 L_2}  ,
\end{align*}
where $\mu_b(\theta)\defeq  
\log \Tr e^{ \log \rho + \theta^1 L_1+\theta^2 L_2} $.  
Hence, we obtain \ding{193}.

Next, we prove that \ding{193}$\Rightarrow$\ding{194}.
We define
$\tilde{\rho}_{\theta}\defeq 
\Pi^{\theta^k}_{X_k,x} 
,\cdots ,\Pi^{\theta^1}_{X_1,b} \rho_{\mix}$
for $\theta=(\theta^1, \ldots, \theta^k)$.
Then, the condition \ding{193} guarantees that
$\tilde{\rho}_{\bar{\theta}}
= \Pi^{1}_{\sum_i (\bar{\theta}^i-\theta^i) X_i,x} 
\tilde{\rho}_{\theta}$.
In particular, when $\theta=0$,
we obtain 
$\tilde{\rho}_{\bar{\theta}}
= \Pi^{1}_{\sum_i \bar{\theta}^i X_i,x} 
\rho_{\mix}$.
Since $\sum_i \bar{\theta}^i X_i$ is commutative with 
$\rho_{\mix}$,
we can apply the classical observation to this case.
Hence, the state $\tilde{\rho}_{\bar{\theta}}$ coincides
with the state $\rho_{\bar{\theta}}$ defined in (\ref{7-3-4}).

Let $\tilde{X}_{j,\theta}$ be the 
$x$-$e$ representation of 
the partial derivative concerning $\theta^j$ at $\rho_\theta$.
It can be expressed as
\begin{align*}
\tilde{X}_{j,\theta}=
X_j - \Tr \rho_\theta X_{j} + \bar{X}_{\theta,j},
\end{align*}
where $\bar{X}_{\theta,j}$ is the skew-Hermitian part.
Thus,
\begin{align*}
&\frac{\partial \Tr \rho_\theta X_j}{\partial \theta^i}
= \Tr \left(\frac{\partial \rho_\theta }{\partial \theta^i}X_j\right)
= \Tr 
\left(\frac{\partial \rho_\theta }{\partial \theta^i}
(X_j -  \Tr \rho_\theta X_{j}) \right)\\
=& \re \Tr 
\left(\frac{\partial \rho_\theta }{\partial \theta^i}
(X_j -  \Tr \rho_\theta X_{j}+ \bar{X}_{\theta,j}) \right)
= \re J_{\theta,x;i,j}.
\end{align*}
Note that the trace of the product of a Hermitian matrix and 
a skew-Hermitian matrix is an imaginary number.
Since $\re J_{\theta,x;i,j} = \re J_{\theta,x;j,i}$,
we have
$\frac{\partial \Tr \rho_\theta X_j}{\partial \theta^i}
= \frac{\partial \Tr \rho_\theta X_i}{\partial \theta^j}$.
Thus, there exists a function $\bar{\mu}(\theta)$
such that $\bar{\mu}(0)= \mu(0)$ and
\begin{align*}
\frac{\partial \bar{\mu}(\theta)}{\partial \theta^i}=
\Tr \rho_\theta X_i.
\end{align*}
This function $\bar{\mu}$ satisfies condition \ding{194}.

Moreover, since $\Tr \rho_{\mix}X_i =0$,
from the definition (\ref{7-3-9}), we have $\bar{\mu}(\theta)
- \bar{\mu}(0)
= D^{\bar{\mu}}(0\|\theta)$.
Since the state $\rho_{\mix}$ commutes the state $\rho_\theta$,
the relation $ D^{(e)}(\rho_{\mix}\|\rho_\theta)= 
\mu(\theta)- \mu(0)$ holds.
Hence, we obtain $\bar{\mu}(\theta)= \mu(\theta)$.

Further, 
we have $D^{\mu}(\bar{\theta}\|\theta)= D(\rho\|\theta)$.
Thus, the equivalence between \ding{194} and \ding{195}
is trivial
since the limit of $D(\rho_{\bar{\theta}}\|\rho_\theta)$
equals the Bogoljubov inner product $J_{b,\theta}$.
Hence, we obtain \ding{195}$\Rightarrow$\ding{192}.

Now, we proceed to the 
proof of 
\ding{192}$+$\ding{193}$+$\ding{194}$+$\ding{195}$\Rightarrow$\ding{196}.
In this case, the function $\nu(\eta)$ coincides with
the Legendre transform of $\mu(\theta)$,
and $\eta_i= \frac{\partial \mu}{\partial \theta^i}(\theta)$.
Hence, $D^{\nu}(\eta\|\bar{\eta})
= D^{\mu}(\bar{\theta}\|\theta)
= D(\rho_{\bar{\eta}}\|\rho_{\eta})$.
The second derivative matrix $\frac{\partial^2 \nu}
{\partial \eta^i \partial \eta^j}$
coincides with
the inverse of the second derivative matrix
$\frac{\partial^2 \mu}
{\partial \theta^i \partial \theta^j}$,
which equals the Bogoljubov 
Fisher information matrix concerning 
the parameter $\theta$.
Since the Bogoljubov Fisher information matrix
concerning the parameter $\eta$
equals the inverse of
the Bogoljubov Fisher information matrix concerning 
the parameter $\theta$,
the Bogoljubov Fisher information matrix
concerning the parameter $\eta$ coincides with
the second derivative matrix $\frac{\partial^2 \nu}
{\partial \eta^i \partial \eta^j}$.
Hence, from (\ref{7-4-4}), we have 
$D^{\nu}(\eta\|\bar{\eta})=
D^{(m)}_b(\rho_{\bar{\theta}}\|\rho_{\theta})$.

Next, we prove \ding{196}$\Rightarrow$\ding{197}.
Since $\rho_{\mix}= \rho_0$ commutes with
$\rho_\eta$,
the $m$ divergence $D^{(m)}_x(\rho_{0}\|\rho_{\eta})$
coincides with 
the Bogoljubov $m$ divergence 
$ D^{(m)}_b(\rho_{0}\|\rho_{\eta})$,
which equals 
the Legendre transform of $\mu(\theta)$
defined in (\ref{7-3-4}).
Thus, $
D^{(m)}_x(\rho_{\bar{\eta}}\|\rho_{\eta})=
D^{\nu}(\eta\|\bar{\eta})=
D(\rho_{\bar{\eta}}\|\rho_{\eta})$.
Finally, taking the limit $\bar{\eta} \to \eta$,
we obtain $J_{x,\eta}= J_{b,\eta}$, {\it i.e.},
\ding{197}$\Rightarrow$\ding{192}.}

\section{Concluding Remark}
In this paper, we 
proved the additivity of $e$-divergences and
the monotonicity of $m$-divergences.
We also found interesting relations between
geometrical path-divergences and an operator-algebraic divergence
as
\begin{align*}
D^{(e)}_r(\rho\|\sigma)=
D^{(m)}_r(\rho\|\sigma)=\overline{D}(\rho\|\sigma) .
\end{align*}
In addition, we obtained the characterization of 
Bogoljubov inner product as Theorem \ref{7-4-5},
which is a generalization of Amari \& Nagaoka \cite{Na}'s characterization.
It is expected that 
these characterizations are applied to quantum information.

\end{document}